# Modeling Link-level Road Traffic Resilience to Extreme Weather Events Using Crowdsourced Data

Songhua Hu [a], Kailai Wang [b*], Lingyao Li [c], Yingrui Zhao [d], Zhenbing He [a], Yunpeng Zhang [e]

a. Senseable City Lab, Massachusetts Institute of Technology, MA, United States

b. Department of Construction Management, University of Houston, TX, United States

c. School of Information, University of Michigan, Ann Arbor, MI, United States

d. Department of Geographical Sciences, University of Maryland, College Park, MD, United States

e. Department of Information and Logistics Technology, University of Houston, TX, United States

***Corresponding authors***: b* kwang43@central.uh.edu, 4230 Martin Luther King Boulevard. #300, Houston, TX 77204-4020, USA

**Abstract**

Climate changes lead to more frequent and intense weather events, posing escalating risks to road traffic. Crowdsourced data offer new opportunities to monitor and investigate changes in road traffic flow during extreme weather. This study utilizes diverse crowdsourced data from mobile devices and the community-driven navigation app, Waze, to examine the impact of three weather events – floods, winter storms, and fog – on road traffic. Three metrics, speed change, event duration, and area under the curve (AUC), are employed to assess link-level traffic change and recovery. In addition, a user-perceived severity is computed to evaluate link-level weather impact based on crowdsourced reports. Results show that overall, winter storms have the greatest impact on road traffic, reducing traffic volume by 58.27% and speed by 12.55%, lasting over 30.87 hours. Floods have a moderate impact, while fog has negligible impact. At a link level, lower-class roads with lower average speeds and volumes experience milder changes during winter storms and floods. Compared to freeways, local streets show 13.73% to 15.21% less speed reduction and require 53.46% to 56.91% shorter time for recovery. Road geometry, such as minimal altitude and slopes, significantly affects road vulnerability under floods but not winter storms. Last, this study reveals that crowdsourced user-perceived severity may not accurately reflect actual link-level impact during severe weather events like winter storms, since few users can travel on the road to serve as "sensors". In sum, this study evaluates a range of new data sources and provides insights into the resilience of road traffic to extreme weather, which are crucial for disaster preparedness, response, and recovery in road transportation systems.

# 1. Introduction

Climate change has led to an increase in the frequency and intensity of extreme weather events worldwide (Stott, 2016), posing significant challenges to the road transportation system. Extreme weather events, such as floods, winter storms, and heavy rainfall/snowfall, all have profound impacts on road traffic (Maze et al., 2006). For example, heavy rainfall reduces driving visibility and pavement friction, leading to lower speed and longer headway (Hu et al., 2018b). Flash floods, mostly caused by heavy rainfall and insufficient drainage capability, lead to lane submersion and reduced roadway capacity (Hu et al., 2018a). Winter storms, characterized by freezing temperature, heavy snowfall, and high wind, result in hazardous road conditions such as icy surface, limited visibility, and lane obstruction or damage (Datla and Sharma, 2008). These weather events not only cause worse traffic status, but also increase the accident risk and jeopardize driver safety (Malin et al., 2019; Markolf et al., 2019). For example, based on 10-year averages from 2007 to 2016 National Center for Statistics and Analysis (NHTSA) data, approximately 21% of vehicle crashes in the US are weather-related (FHWA, 2023). These circumstances highlight the need for understanding weather impact and developing effective strategies to ensure infrastructure functionality under diverse weather conditions.

Road traffic resilience, which assesses the ability of road traffic to withstand and recover from weather disturbances, has gained increasing attention in recent decades (Balakrishnan et al., 2020; Qiang and Xu, 2020). However, most previous studies heavily rely on data from static detectors and field-based observations (Datla and Sharma, 2008; Knapp and Smithson, 2000; Maze et al., 2006), which require significant effort and resources for pre-deployment and data compilation (Hu, 2023). In recent years, crowdsourcing has emerged as a promising and cost-effective alternative for gathering large-scale road, traffic, and weather information, either through passively-collected spatiotemporal footprint or user-posted social media content (Yabe et al., 2022). Representatives include using floating cars to estimate traffic speed and volume (Hu et al., 2018b), leveraging mobile device locations to estimate travel demand (Hu et al., 2023a), and utilizing community-driven navigation apps (e.g., Waze) to collect traffic incidents and weather events (Lenkei, 2018; Li et al., 2021a; Praharaj et al., 2021b).

Despite the significant advancements that crowdsourced data has brought to road traffic resilience analysis, most studies have primarily focused on comparing the spatiotemporal consistency between crowdsourced data and field-collected data to evaluate data trustworthiness (Amin-Naseri et al., 2018; Lenkei, 2018; Lowrie et al., 2022). Yet, a comprehensive framework that integrates different data sources, models their relationships, constructs link-level traffic resilience measures, and identifies key factors related to resilience remains lacking. In addition, most studies have focused on individual weather events with limited spatiotemporal coverage (Akin et al., 2011; Lowrie et al., 2022), and only a few have delved into the link level to understand traffic resilience (Praharaj et al., 2021b), leaving a gap in understanding how the impact varies across different weather events or road segments with different characteristics. To address these research gaps, three key questions are proposed:

1) How does road traffic resilience vary across different weather events?
2) Can crowdsourced data be considered a reliable source for modeling road traffic resilience during extreme weather?
3) What are the key factors related to road link-level traffic resilience?

Specifically, this research utilizes multi-source data from static detectors, local agencies, and crowdsourced datasets to examine the impact of three weather events – flash floods, winter storms, and fog – on road traffic dynamics. Focusing on the Dallas-Fort Worth (DTW) area in 2022, this study first validates data reliability by comparing traditional data with crowdsourced data. Then, three resilience metrics – speed change, event duration, and area under the curve (AUC) – are computed to assess link-level traffic resilience; meanwhile, a user-perceived event severity is constructed based on crowdsourced reports from Waze to assess the link-level weather impact. Last, a set of generalized additive regressions are fitted to understand relationships among road characteristics, user-perceived severity, and traffic resilience. This study evaluates a range of new datasets and provides versatile methods to measure road traffic resilience. Findings shed light on improving preparedness plans, enhancing road maintenance strategies, and bolstering overall road safety during extreme weather.

## 2. Literature review

### 2.1 Impact of extreme weather on road traffic

Extreme weather events mainly refer to unusual weather or climate conditions that can cause devastating impacts on communities and natural ecosystems, such as freezes, hails, fogs, heavy rains, snows, tornadoes, and floods (Koetse and Rietveld, 2009). Their effects on road traffic have been extensively studied. Generally, extreme weather significantly reduces travel demand (Datla and Sharma, 2008; He et al., 2021; Zhou et al., 2017), impairs network accessibility (Esfeh et al., 2022; Shen and Kim, 2020; Zhang and Alipour, 2021), worsens traffic conditions (e.g., decreased volume, speed, and capacity, increased delay) (Bi et al., 2022; Hu et al., 2018a; Zhao and Zhang, 2020), and increases the risk of incidents (Theofilatos and Yannis, 2014).

Although the impacts of adverse weather on road traffic consistently yield negative effects, their magnitudes vary significantly across event types, weather intensity levels, road types, and space and time. Early studies summarized that the reduction in freeway speed and capacity increases with higher intensity levels of rain, snow, fog, and wind (Maze et al., 2006). Their results showed that snow leads to the greatest reduction, followed by rain and fog. Similar findings are documented in subsequent studies, including the Highway Capacity Manual (HCM) 2010 (Akin et al., 2011; Hu et al., 2018b; TRB, 2010). Weather impacts also differ across road types. For example, commuter roads experience lower reductions in traffic volume due to winter snow compared to recreational roads (Datla and Sharma, 2008); freeways experience the greatest reductions in traffic volume due to recurring floods compared to other road types (Praharaj et al., 2021a). Last, weather impacts vary significantly across regions and times. For example, one study compared 64 winter storm events across 7 U.S. states and found volume reductions ranged from 16% to 47% (Knapp and Smithson, 2000). Several studies also documented differences in weather impacts on traffic between peak hours and non-peak hours, daytime and nighttime, and weekdays and weekends (Bi et al., 2022; Hu et al., 2018b; Koetse and Rietveld, 2009).

## 2.2 Resilience of road traffic network

Recently, a shift has been observed from simply examining the impact of weather events on road traffic networks to exploring their resilience (Markolf et al., 2019). In a traffic network, resilience refers to the ability of the network to resist (aka vulnerability) and recover from disturbances (Calvert and Snelder, 2018). Studies related to road traffic resilience can be categorized into two distinct traditions (Mattsson and Jenelius, 2015; Pan et al., 2021). The first tradition centers on predicting the spatiotemporal dynamics and identifying the most vulnerable (or critical) nodes or links in a network before a hazard occurs. Related approaches include hydrodynamic simulation, meteorological models, traffic simulation, graph theory, deep learning, or a combination of them (Kasmalkar et al., 2020; Rahman and Hasan, 2023; Singh et al., 2018; Wang et al., 2020a; Zhang and Alipour, 2020, 2021).

Another tradition focuses on quantifying road traffic resilience during and after disruptions and identifying relevant key factors using data-driven methods (Chen et al., 2023; Pan et al., 2021). Measures such as the magnitude of impact, time to recovery, or a combination of both, are commonly used to quantify road traffic resilience (Hong et al., 2021; Qiang and Xu, 2020). Road attributes (e.g., road functional class, number of lanes, topology, geometry), traffic flow conditions (e.g., average speed, volume), and community features (e.g., surrounding socioeconomic status, land use types), are widely recognized as key determinants of road traffic resilience (Balakrishnan et al., 2020; Chen et al., 2023; Shen and Kim, 2020). For example, high-class roads located in urbanized areas have been observed to experience greater impacts during adverse weather conditions (Malin et al., 2019; Praharaj et al., 2021a). Socially disadvantaged communities are found to suffer the largest impacts and face the greatest challenges in post-event recovery (Chen et al., 2023; Hu et al., 2023b; Shen and Kim, 2020).

## 2.3 Crowdsourced data in understanding weather impacts

Conventional research on the impact of extreme weather on road traffic typically relies on data collected from roadside detectors and weather stations (Akin et al., 2011). However, these data collection methods are time-consuming and resource-intensive (Chen et al., 2019; Hu, 2023). In contrast, crowdsourcing has emerged as a promising and cost-effective alternative for gathering near-real-time large-scale human travel information (Hu, 2023; Hu et al., 2023a). Literature has demonstrated the superiority of crowdsourcing in monitoring human mobility during extreme weather events, making it a valuable complement or even replacement for traditional sensors and surveys (Li et al., 2021a; Yabe et al., 2022; Zhou et al., 2017).

One of the most popular types of crowdsourced data in road traffic analysis is floating car data, which has been used for estimating traffic speed, volume, and density (Hu et al., 2018a). These traffic flow measures are integrated into navigation maps to represent real-time traffic congestion or serve as main products for several established commercial data companies like INRIX, TomTom, and Streetlight (Xing et al., 2022). Social media also offers an alternative for crowdsourcing traffic information (Li et al., 2021b). With vast user-shared content such as messages and images, social media provides a unique opportunity for rapidly locating road damage or incidents (Hu et al., 2018b).

Recently, Waze has emerged as a promising crowdsourcing tool in road traffic research. Waze combines social elements into its navigation map, encouraging users to share the time and location of traffic incidents or weather events on the road. This enables the easy and real-time gathering of accurate and large-scale road traffic event data. While prior studies primarily focused on road safety using Waze data (Amin-Naseri et al., 2018; Lenkei, 2018), its application in road traffic resilience during weather events is gaining momentum. Recent studies have leveraged Waze to locate flash floods (Lowrie et al., 2022) on road networks, investigate the impacts of extreme weather on road traffic volume (Praharaj et al., 2021a; Praharaj et al., 2021b), and predict road flooding risk by combining Waze data with traffic data (Yuan et al., 2023).

## 3. Research design

In this study, multi-type data sources from static detectors, local agencies, and crowdsourced datasets are used to investigate traffic flow changes during different weather events. The research framework is illustrated in **Figure 1**. First, crowdsourced data are compared with agency-reported data to demonstrate whether they can well capture weather impact on traffic flow. Then, all data are spatiotemporally integrated with the road network to quantify both network-level and link-level traffic resilience and user-perceived severity. Last, by considering road link characteristics and user-perceived severity as dependent variables, this study explores the key factor related to the impact of weather events on traffic flow status.

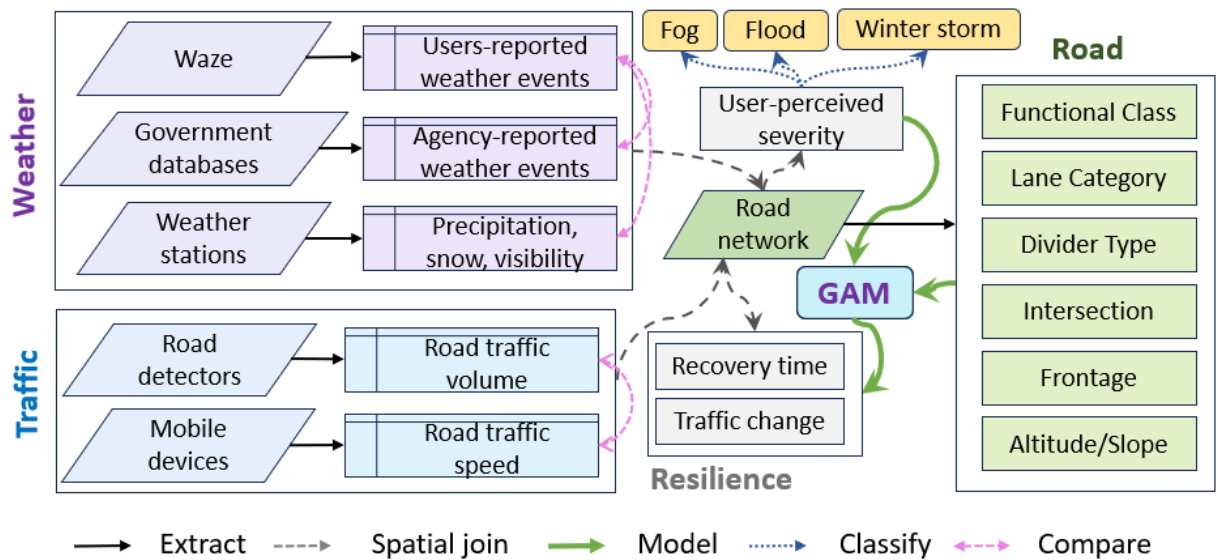

**Figure 1**. Research framework.

### 3.1 Study area

The Dallas–Fort Worth (DFW) metroplex in North Texas serves as the case scenario in this study due to its long history of diverse weather events. With a humid subtropical climate, DFW is characterized by hot and humid summers and mild winters. The region is particularly vulnerable to flash floods during the summer due to its low-lying areas and susceptibility to heavy rainfall. These localized floods can lead to road closures, traffic disruptions, and safety risks for drivers.

An illustrative event occurred on August 22, 2022, when thunderstorms hit the area, resulting in inundated streets, flooded homes, damaged properties, and tragic fatalities.

During the winter season, DFW encounters cold temperatures, occasionally dipping below freezing. Although winter storms are infrequent in DFW, they can cause significant impacts on road infrastructure, transportation services, power supply, and daily activities, as DFW is not as well-prepared for winter weather as more northern areas. The disastrous 2021 winter storm stands as a poignant example, crippling the Texas power grid, claiming numerous lives, and leaving millions without heat for days (Chen et al., 2023). In February 2022, DFW experienced several winter storms that, while not as severe as the 2021 storm, still had a notable impact on the region. These storms resulted in icy roads, freezing temperatures, and widespread closures of schools, roads, communities, and transportation services (Dey and Douglas, 2022).

The diverse range of weather events in DFW, coupled with the abundance of available data in this area, makes it an ideal scenario for analyzing and comparing the impact of various weather events on urban road traffic. Findings can provide valuable insights and actionable information to tackle the challenges posed by weather disturbances and enhance the overall preparedness of the area.

**3.2 Data preparation**

**3.2.1 Weather information**

***Crowdsourced weather event dataset (Waze)***: Waze is a community-driven mobile navigation application that incorporates a real-time information reporting tool within its navigation map, enabling users to conveniently report various traffic events on the road. These events encompass typical road incidents such as road construction, lane closures, accidents, as well as diverse weather events including fog, flood, icy road, hail, heavy rain/snow, etc. Waze promptly displays these reports on the map in real-time, providing travelers with up-to-date information on potential delays and congestion. Each report is documented with start time, end time, precise location coordinates, road name, and direction. Additionally, a reliability score is assigned to evaluate the reliability of each report. For this study, weather-related reports from Waze within the DFW area during the year 2022 were extracted via the Waze for Cities data-sharing program. A total of 7,244 reports were received, with 51.82% attributed to floods (recorded as "Flood" or "Heavy Rain"), 16.44% to winter storms (recorded as "Road Icy" or "Heavy Snow"), 27.82% to fog, and 3.92% to other events. Consequently, the top three weather events - floods, winter storms, and fog - were selected as the research target.

***National Centers for Environmental Information (NCEI) Storm Events Dataset***: The NCEI Storm Events Database contains the records used to generate the official National Oceanic and Atmospheric Administration (NOAA) Storm Data publication. It documents three main types of weather events: 1) Significant weather phenomena having sufficient intensity to cause loss of life, injuries, property damage, and/or disruption to commerce; 2) Rare weather phenomena that generate media attention; and 3) Other significant meteorological events, such as record maximum or minimum temperatures or precipitation. For this study, weather events related to floods (recorded as "Flash Flood" and "Flood") and winter storms (recorded as "Winter Storm") occurring in the DFW area during the year 2022 were extracted as the baseline for comparing with the crowdsourced data. Note that fog is not considered a distinct weather event by NOAA, and thus, no related records on fog could be obtained.

*Weather station dataset*: The DFW area is equipped with 12 static weather stations that collect detailed meteorological information, including precipitation, snow cover, temperature, wind speed, visibility, etc. The data from these stations are aggregated on an hourly basis and archived by NOAA NCEI. For this study, the focus lies on precipitation, snow cover, and visibility, which correspond to the three main user-reported weather events extracted from Waze.

### 3.2.2 Road traffic information

*Texas Department of Transportation (TxDOT) traffic volume data*: Most of the traffic stations in the DFW area collect snapshot traffic count data to produce Annual Average Daily Traffic (AADT). Only 15 permanent continuous counters in this area gather road traffic volume 24 hours a day, 365 days a year (**Figure 2** (a)). For this study, the hourly traffic volume from these 15 permanent counters in the year 2022 is extracted from the Statewide Traffic Analysis and Reporting System (STARS II). Although the limited number of traffic counters cannot cover the entire DFW metroplex adequately, it can serve as an average metric for analyzing network-level temporal changes in traffic volume under different weather events.

*Crowdsourced traffic speed data (INRIX)*: INRIX is a location-based services company that collects aggregated, anonymous location data from 500 million vehicles and mobile devices worldwide in real time (Xing et al., 2022). It covers both consumer and fleet vehicles to calculate high-resolution link-level traffic speed and travel time. For this study, the hourly link-level traffic speed for 19,179 road segments in the DFW area in the year 2022 are extracted. Although not all road links are covered by the INRIX speed dataset, the spatial coverage of crowdsourced speed is considerably wider compared to the volume obtained from permanent counters (see **Figure 2** (a)), making it a reliable source for capturing network-representative speed changes.

***Road network (HERE)*:** While INRIX focuses on speed, it lacks link-level road characteristics. HERE Map offers navigable road networks with extensive link attributes and comprehensive coverage. Compared with Open Street Map, HERE Map offers more extensive link attributes, such as functional classes, speed rank, number of lanes, divider type, intersection, geometry, etc. These attributes are valuable for analyzing how traffic status varies across different road characteristics during extreme weather. Note that the HERE Map also offers additional labels to indicate whether the road is a ramp, bridge, tunnel, etc. However, due to the limited availability of speed data for these roads, this study excludes them from the analysis.

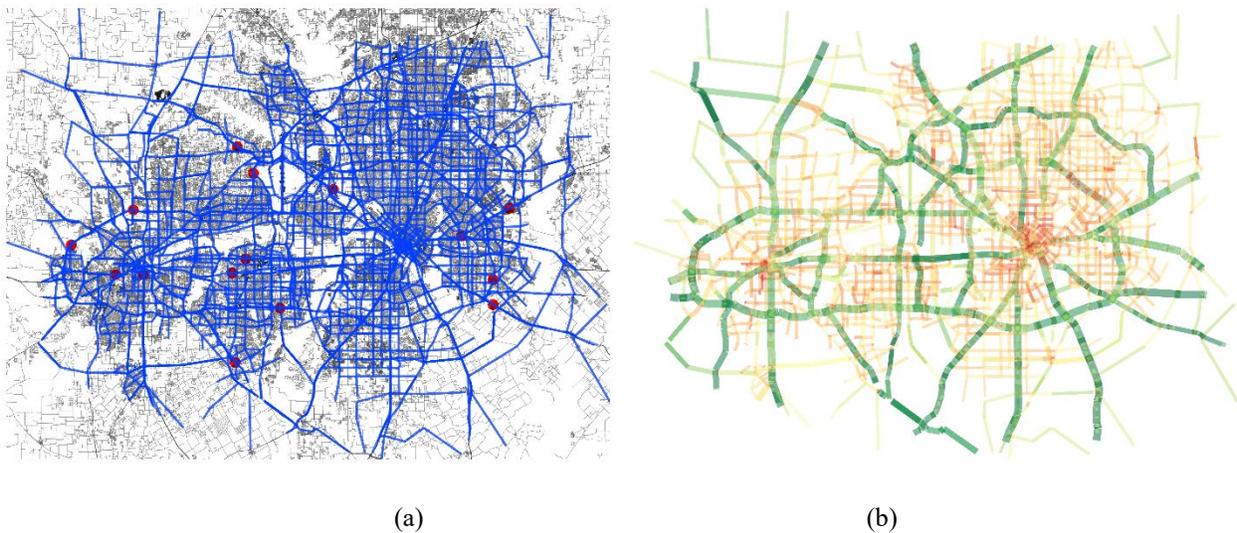

(a) (b)

**Figure 2**. The study area. (a) Road networks. Gray lines depict the comprehensive road network from HERE Map. Blue lines represent the road network with available speed data from INRIX. Red points indicate permanent continuous traffic counters. (b) Road networks by functional class (from HERE Map) and average speed (from INRIX). Wider lines denote higher functional class, and greener lines indicate faster average speed.

### 3.2.3 Data fusion

One main challenge in this study is the fusion of diverse data sources into a unified dataset. Waze user-reported weather events, represented as point data, need to be integrated with the HERE Map. To achieve this, we established three merging rules: the distance between the event location and the link should be less than 10 meters, the directions should match, and the road names should be consistent. Eventually, 79.54% of the events are successfully attached to their corresponding road segments. Additionally, we merged the HERE Map with the INRIX speed data using the Traffic Message Channel (TMC) code. It is important to note that each TMC code may represent multiple links in the HERE Map. To address this, we computed the mode for categorical link attributes and the average for numerical features, weighted by the link length. The resulting merged network is mapped in **Figure 2** (b), where higher-class roads are associated with higher average speeds, validating the accuracy of our fused network.

### 3.3 Variable description

In this study, the resilience of link-level road traffic is assessed using three metrics: the extent of speed change, the duration of the change, and the accumulated change of speed over the entire process. Since other traffic flow characteristics are unavailable at the link level, speed is used as the sole metric for resilience computation. These three metrics are then employed as dependent variables in subsequent regression analyses.

    **Figure 3** provides a visual representation of the three metrics. First, the average speed during normal periods is computed for each link, categorized by hour of the day and day of the week. Then, the relative speed change is computed by comparing the observed speed with the average normal speed (represented by the bold curves in **Figure 3**). Second, a threshold, set at -

1% (illustrated by dashed lines in **Figure 3**), is employed to determine the start and end of the event. Only when the relative change falls below -1% is considered as being affected. Third, the maximum reduction in speed during the event is used as an indicator of the extent of speed change. Last, similar to (Hong et al., 2021; Qiang and Xu, 2020), the area under the curve (AUC) (depicted by the green area in **Figure 3**) is integrated to represent the accumulated speed change based on the composite trapezoidal rule, which approximates the integral by evaluating the integrand at multiple points without requiring knowledge of the function form. Let $a$ represent the event start time and $b$ represent the end, with $[a, b]$ denoting the integration interval with a partition $a = t_0 < t_1 < \ldots < t_n = b$, then the three metrics can be expressed as follows:

$$\text{Duration} = |b - a| \quad (1)$$

$$\text{Change} = \min\{f(t) = \frac{s(t) - \bar{s}(t)}{\bar{s}(t)} : a \leq t \leq b\} \quad (2)$$

$$\text{AUC} = \int_a^b f(t)dt \approx \frac{1}{2}\sum_{j=1}^{n}(t_j - t_{j-1})[f(t_{j-1}) + f(t_j)] \quad (3)$$

where $f(t)$ is the time function of relative speed change; $s(t)$ is the time function of observed speed; $\bar{s}(t)$ is the time function of average speed, which is computed using historical speed from normal periods within the last month prior to the occurrence of the event.

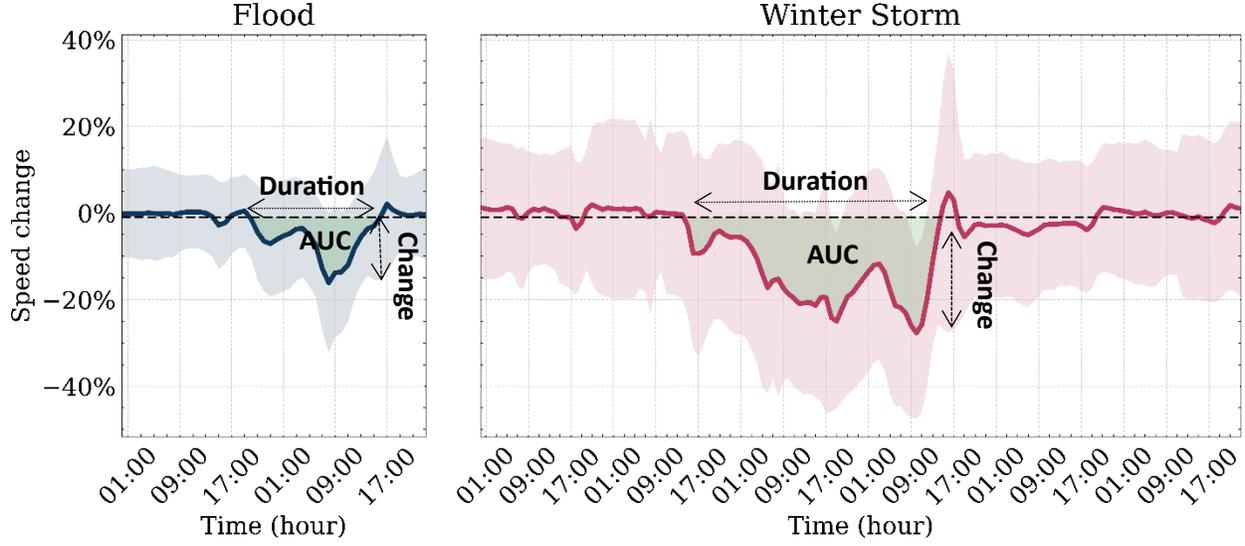

**Figure 3**. Illustration of three resilience metrics. Bold lines are the speed change relative to the normal speed observed during normal periods. Buffers are the 95% confidence interval.

The independent variables primarily consist of various link attributes, including functional class, lane category, divider type, intersection, frontage, and geometry (see **Table 1** for the summary). Additionally, a user-perceived severity is constructed based on reports from Waze to measure users' perceptions of weather impact on each link. The underlying assumption is that if a particular link receives a higher density of highly reliable reports, it is more likely to be significantly impacted by the weather event:

$$\text{UPS} = \sum_{i=1}^{m} \frac{w_i}{10L} \qquad (4)$$

where UPS is the user-perceived severity for one link; $w_i$ is the reliability score of the $i^\text{th}$ report related to the weather event on that link, with $m$ being the number of reports; $L$ is the link length. Note that $w_i$ is directly assigned by Waze with a value between 0 and 10, with higher scores indicating greater reliability. The assigned score takes into account user reactions, such as "Thumbs up" or "Not there," as well as the level of expertise of the reporting user.

Variable selection was performed to determine the optimal variable set. The variance inflation factor (VIF) was first calculated to test the multicollinearity, and VIFs greater than 5 were excluded. Then, a forward stepwise regression was used to select the optimal independent variables based on the smallest AIC. Note that the average speed and AADT are both excluded from the model due to their high correlation with road characteristics like road functional classes and lane categories. To understand their effects, we build a separate model by only considering average speed and AADT as independent variables.

Table 1. Summary statistics (link-level)

| Variable | Description | N | Mean | Std. |
|---|---|---|---|---|
| **Dependent Variables** | | | | |
| **Flood (August 21 - 23, 2022)** | | | | |
| Duration | The duration of speed change, in hours. | 19179 | 11.20 | 5.78 |
| Speed change | The maximum speed reduction, in %. | 19179 | -16.39 | 16.01 |
| AUC | The accumulated speed change, in hours·%. | 19179 | -144.50 | 123.20 |
| **Winter Storm (February 2 - 5, 2022)** | | | | |
| Duration | The duration of speed change, in hours. | 19179 | 30.87 | 20.03 |
| Speed change | The maximum speed reduction, in %. | 19179 | -24.41 | 18.76 |
| AUC | The accumulated speed change, in hours·%. | 19179 | -742.60 | 574.90 |
| **Independent Variables** | | | | |
| **Functional Class** | | | | |
| Freeway (Reference) | Roads allow for high volume, maximum speed traffic movement between and through cities/metropolitan areas. | 3866 | 20.16% | |
| Arterial | Roads that connect freeways and provide a high volume at a lower level of mobility than freeways. | 2043 | 10.65% | |
| Collector | Roads provide for a high volume of traffic movement at moderate speeds between neighborhoods. | 12211 | 63.67% | |
| Local street | Roads whose volume is below any functional class, such as walkways, truck-only roads, access roads, and parking lanes. | 1059 | 5.52% | |
| **Lane Category** | | | | |
| 1 lane (Reference) | Roads with one Lane. | 3971 | 20.70% | |
| 2-3 lanes | Roads with two or three Lanes. | 13930 | 72.63% | |
| >3 lanes | Roads with four or more Lanes. | 1278 | 6.66% | |
| **Divider Type** | | | | |
| No divider (Reference) | Roads with no divider. | 18736 | 97.69% | |
| Legal divider | Roads with legal dividers can be traversed. | 181 | 0.94% | |
| Physical divider | Roads with physical dividers that cannot be traversed. | 262 | 1.37% | |
| **Intersection** | | | | |
| Not an intersection (Reference) | Roads that do not belong to an intersection. | 12645 | 65.93% | |
| Intersection | Roads that are intersection internal link. | 6534 | 34.07% | |
| **Frontage** | | | | |

| Not a frontage (Reference) | Not a frontage road. | 17873 | 93.19% | |
|---|---|---|---|---|
| Frontage | Frontage roads that run parallel to a road with a higher traffic flow (aka Service Roads). | 1306 | 6.81% | |
| **Geometry** | | | | |
| Min. altitude | The minimum height along the road, in km. | 19179 | 14.41 | 2.86 |
| Slope | The slope of the road. | 19179 | 1.47 | 1.34 |
| **Normal traffic flow** | | | | |
| Speed | Average speed during normal periods, in mph. | 19179 | 34.8 | 16.29 |
| AADT | Annual average daily traffic, in count. | 8042[a] | 60609 | 55482 |
| **User-perceived severity** | | | | |
| Flood severity | The density of flood events reported by Waze users along roads, weighted by the reliability score, in count/mile. | 19179 | 0.15 | 1.77 |
| Snow severity | The density of snow events reported by Waze users along roads, weighted by the reliability score, in count/mile. | 19179 | 0.02 | 0.31 |

a. AADT is from the Highway Performance Network provided by the US Department Of Transportation Federal Highway Administration. Only 8042 links have AADT data.

## 3.4 Model: Generalized additive model (GAM)

Two main challenges should be addressed in fitting the regressions. First, the distributions of dependent variables are different, some of them may not meet the normality assumption. Second, since the link-level traffic flow is spatially dependent, the spatial autocorrelation should be addressed when fitting models. To this end, we employed the generalized additive model (GAM). GAM (Wood, 2003) is a semi-parametric generalized model that incorporates a linear predictor involving a series of additive non-parametric smooth splines of covariates (Wood, 2003). By utilizing different spline functions and link functions, GAM can accommodate various distributions and nonlinear effects within a unified framework (Wang et al., 2020b):

$$g(E(Y)) = \beta_0 + \sum_{k=1}^{N_L} \beta_k X_k + \sum_{k=1}^{N_N} f_z^{(k)}(Z_k) + s_{i,j} + \varepsilon \quad (5)$$

where $Y$ is one of the six dependent variables, among which speed change and AUC are assumed to follow Gaussian distributions, while duration is assumed to follow a negative binomial (NB) distribution; $g(.)$ is the link function, with a logarithmic form for NB distribution and an identity form for Gaussian distribution; $\beta_0$ is the overall intercept; $\beta_k$ is the coefficient of the $k^{th}$ variable $X_k$ and $N_L$ is the number of variables with linear effects; here we standardized all numeric variables by subtracting their mean and dividing by two times their standard deviation while leaving categorical variables unchanged (Gelman, 2008); $f_z^{(k)}(.)$ is the spline function of the $k^{th}$ variable $Z_k$ and $N_N$ is the number of variables with nonlinear effects; $s_{i,j}$ is an nonlinear coordinate interactive term used to account for spatial autocorrelation; $\varepsilon$ is the error term capturing the unexplained variation.

## 4. Results

### 4.1 Comparison between crowdsourced and traditional data

This section offers a comprehensive comparison between crowdsourced and traditional data. **Figure 4** illustrates the consistent temporal patterns observed across all data sources. The major weather events, such as the August flood and the two February winter storms, are effectively captured by all data. However, crowdsourced data contain more weather events compared to agency-reported data, with many of them aligning with weather conditions monitored by weather stations. For example, Waze users reported road floods on 96 days with heavy rain in 2020, whereas NCEI labeled only 4 of them as floods. In addition, within each event, crowdsourced data contain substantially more reports than agency records. During the August flood, Waze users contributed 3754 reports distributed across the road network, while NCEI collected only 27 reports (**Figure 5** (a) vs. (b)). Similarly, Waze users reported 1191 instances of icy roads (**Figure 5** (c)) during the first winter storm in February 2022, while only 2 reports are collected by the NCEI. In sum, the extensive spatiotemporal coverage of crowdsourced data is a significant advantage compared to traditional agency sources.

Another noteworthy difference is that crowdsourced data tend to detect events earlier than official agencies and document longer-lasting durations of impact. As shown in **Table 2**, the duration captured by Waze users is typically 1.5 to 3 times longer than the NCEI announced duration. This indicates although the official weather alerts have been lifted due to the cessation of rainfall or snowfall, the road network continues to experience lingering effects that require additional time for recovery. In this regard, crowdsourced data provide a new and more accurate way to determine the event's start and end based on users' actual perceptions.

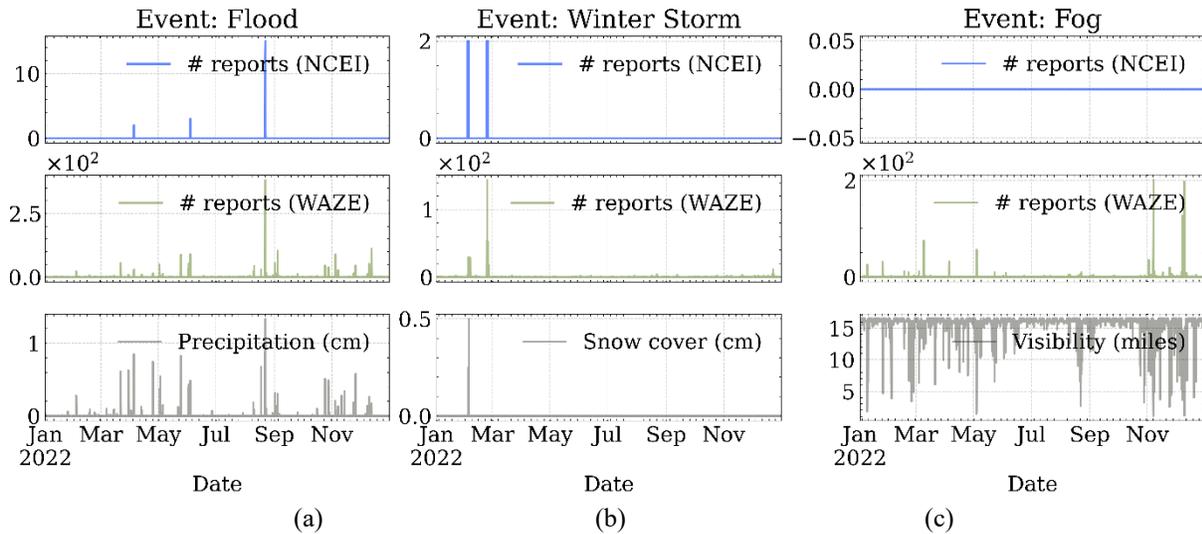

**Figure 4**. Time series of the number of reports from NCEI (the first row) and Waze (the second row), and time series of weather conditions recorded by weather stations (the third row) for three main weather events - Flood, Winter Storm, and Fog - occurring in the DFW metroplex in 2022.

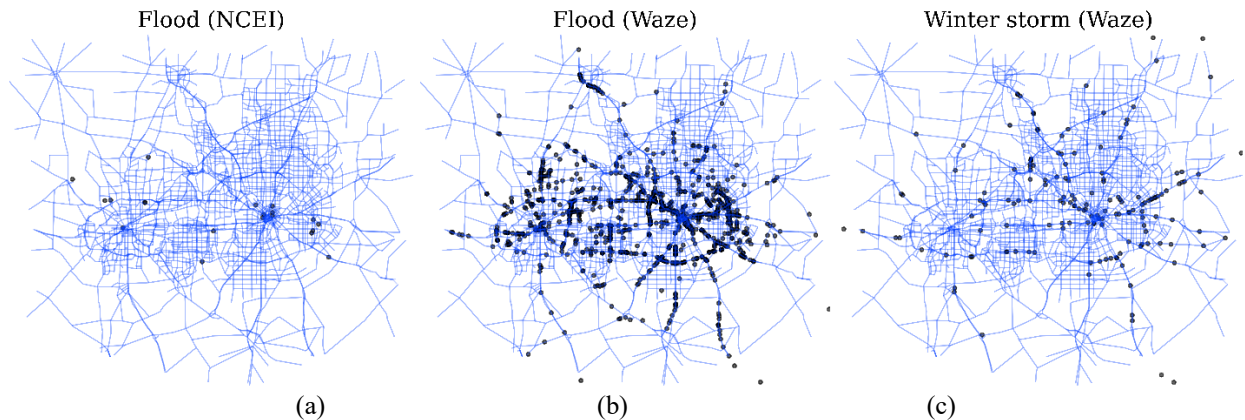

(a)                  (b)                  (c)

**Figure 5**. Spatial map of reports in the DFW area. (a) Flood (August 21 - 23, 2022) (NCEI); (b) Flood (August 21 - 23, 2022) (Waze); (c) Winter Storm (February 2 - 5, 2022) (Waze).

**Table 2**. Duration and number of reports from different data sources

|  | NCEI | Waze |
|---|---|---|
| **Flood in August 2022** | | |
| No. of reports | 27 | 3754 |
| Time interval | 2022-08-21 22:00 to 2022-08-22 21:00 | 2022-08-21 15:00 to 2022-08-23 20:00 |
| Duration (hour) | 23 | 53 |
| **First winter storm in February 2022** | | |
| No. of reports | 2 | 1191 |
| Time interval | 2022-02-02 18:00 to 2022-02-03 18:00 | 2022-02-02 16:00 to 2022-02-05 16:00 |
| Duration (hour) | 24 | 72 |
| **Second winter storm in February 2022** | | |
| No. of reports | 2 | 1507 |
| Time interval | 2022-02-23 02:00 to 2022-02-24 12:00 | 2022-02-23 05:00 to 2022-02-25 10:00 |
| Duration (hour) | 34 | 53 |

**4.2 Impacts of main weather events on traffic flow**

This section summarizes the impacts of main weather events on traffic flow characteristics. Four distinct periods with significantly higher numbers of Waze users' reports are zoomed in and shown in **Figure 6**: the first two columns depict the two February winter storms, the third column represents the August flood, and the last column represents the December fog. As shown, the most drastic decrease in both volume and speed is observed during winter storms, followed by floods, while the impact of fog is relatively inconspicuous. Another visible pattern is that the duration of winter storm impact is much longer than that of floods, which aligns with the average recovery duration reported in **Table 1** and **Table 2**: on average, road traffic takes approximately 11.20 hours to recover from the flood and requires 30.87 hours to recover from the winter storm.

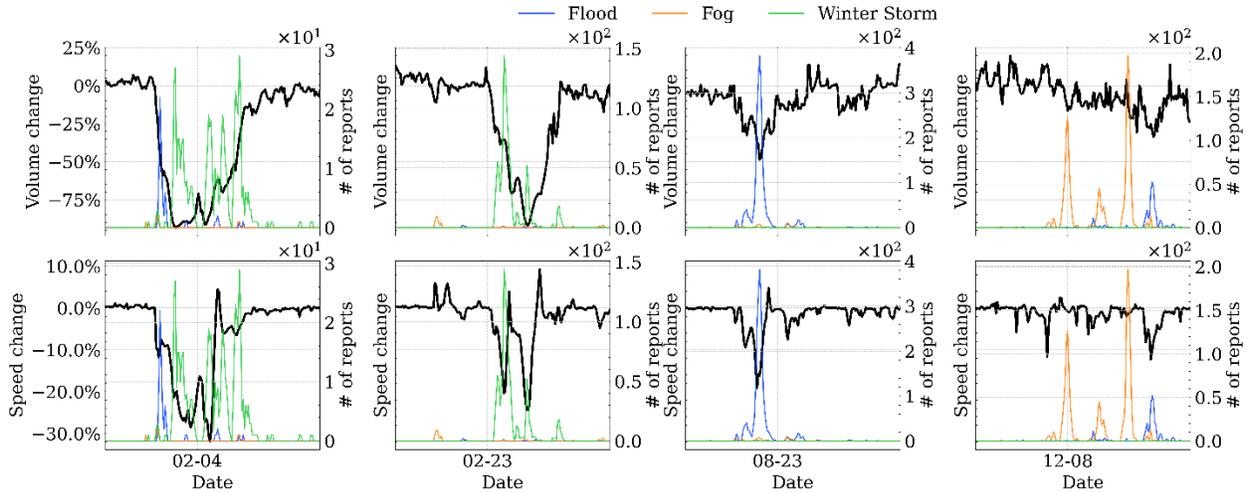

**Figure 6**. Time series of road traffic volume change (top row) and speed change (bottom row) during four main weather events in the DFW metroplex in 2022. The bold black curves (on the left y-axis) represent the relative changes in traffic flow characteristics compared to normal conditions. The blue, orange, and green curves (on the right y-axis) represent the number of events reported by Waze users related to floods, fog, and winter storms, respectively.

To quantify the difference, an unpaired t-test is employed to compare the mean value of hourly speed and volume among normal, minor, and heavy weather events, as shown in **Table 3**. Results reveal that at a network level, the weather impact intensifies as the weather level increases. Different weather events also exhibit varying impacts. Heavy winter storms have the most significant impact, leading to a 58.27% decrease in volume and a 12.55% decrease in speed, followed by heavy floods, leading to a 14.13% decrease in volume and a 4.60% decrease in speed. The impact of fog on traffic is limited; hence, it is excluded from the subsequent modeling process. Another noteworthy finding is that the weather impact on volume is considerably greater than that speed. However, one caveat here is that volume data come from 15 continuous traffic counters, primarily installed on high-class roads like freeways; thus, the volume change reported in this study is not as representative as the speed change obtained from network-wide crowdsourced data.

**Table 3**. T-test of traffic status between normal periods and different disasters

| Type | Level | Volume change (%) | Difference in mean | Speed change (%) | Difference in mean |
|---|---|---|---|---|---|
| Flood | None | -1.82 (11.70) | -- | 0.03 (1.15) | -- |
| | Light | -6.66 (16.97) | -4.84*** (-6.39, -3.29) | -1.08 (3.33) | -1.11*** (-1.41, -0.81) |
| | Heavy | -15.95 (20.33) | -14.13*** (-17.40, -10.86) | -4.57 (4.40) | -4.60*** (-5.30, -3.89) |
| Fog | None | -1.82 (11.70) | -- | 0.03 (1.15) | -- |
| | Light | -0.29 (14.30) | 1.54. (-0.14, 3.21) | -0.29 (1.94) | -0.31** (-0.54, -0.09) |
| | Heavy | -4.78 (8.88) | -2.96* (-5.67, -0.25) | -0.52 (1.03) | -0.55** (-0.86, -0.23) |
| Winter storm | None | -1.82 (11.70) | -- | 0.03 (1.15) | -- |
| | Light | -21.67 (28.97) | -19.85*** (-23.46, -16.24) | -2.36 (7.50) | -2.39*** (-3.32, -1.45) |
| | Heavy | -60.09 (25.47) | -58.27*** (-65.76, -50.79) | -12.52 (10.34) | -12.55*** (-15.59, -9.51) |

Note: Light weather events are defined as the number of hourly Waze reports in the DFW area between 1 and 10, while heavy weather events are defined as the number of related reports exceeding 10.

The spatial distributions of dependent variables are shown in **Figure 7**. As shown, speed change and AUC have similar patterns, while they exhibit an inverse pattern to event duration. This is plausible since more negative speed change, more negative AUC, and longer event duration are all indicative of more severe weather impacts. Another pattern is observed in the comparison between high-class roads (wider links in **Figure 7**) and local roads. High-class roads generally demonstrate greater speed reduction, longer duration, and more negative AUC, suggesting significant differences in weather impacts on traffic status across road types.

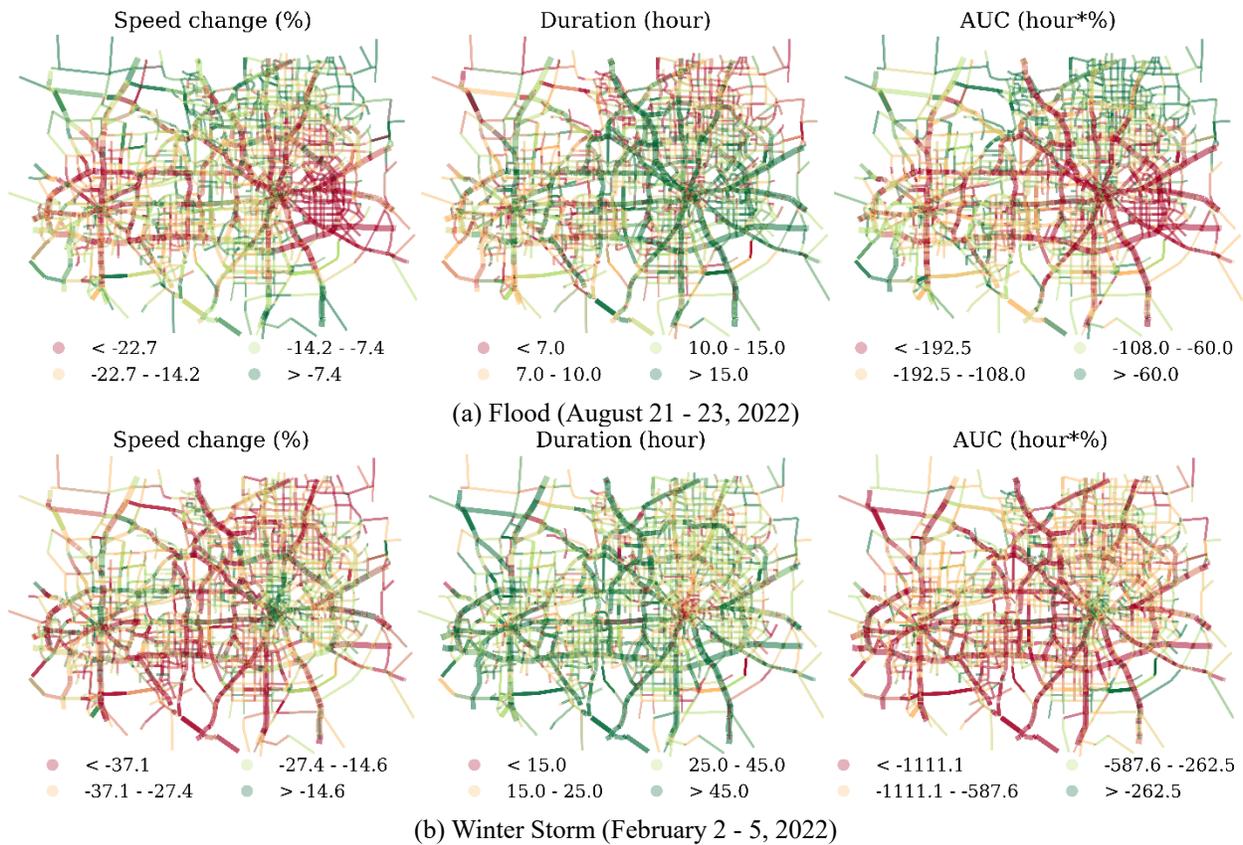

**Figure 7.** Spatial distribution of three dependent variables: traffic speed change (the first column), event duration (the second column), and AUC (the third column). The top row represents the August flood while the bottom row represents the February first winter storm. Link width represents the functional class. Wider links correspond to a higher class. Link color represents the magnitude of the variable.

## 4.3 Relationship between traffic resilience and road characteristics

The results of six cross-sectional GAMs are summarized in **Table 4**. The models consist of two parts: the parametric coefficients representing linear fixed effects and the nonparametric smooth terms representing nonlinear effects. In this study, the effects of most variables are considered linear since they are categorical. Only the spatial interaction terms are treated as nonlinear, and their estimated degrees of freedom (e.d.f.) are all significantly greater than 1, implying that the GAMs well captured the spatial nonlinear dependence. The goodness-of-fit, *Deviance explained*, which represents the proportion of the null deviance explained by the model, ranges from 12.2% to 39.8%, suggesting the independent variables can moderately explain road traffic resilience. The model's goodness-of-fit for AUC is better than for the other two measures. This might be attributed to AUC measuring resilience more comprehensively and thus is more predictable.

Table 4. Model estimation outcomes (Standardized).

| Flood (August 21 - 23, 2022) | | | |
|---|---|---|---|
| Variables | Duration | Speed Change | AUC |
| (Intercept) | 2.702*** (2.687, 2.716) | -21.116*** (-21.848, -20.385) | -236.982*** (-241.939, -232.025) |
| **Linear effects (Parametric coefficients)** | | | |
| Arterial | -0.259*** (-0.275, -0.243) | 5.819*** (4.975, 6.664) | 94.341*** (88.612, 100.070) |
| Collector | -0.458*** (-0.470, -0.446) | 8.989*** (8.358, 9.620) | 135.935*** (131.656, 140.213) |
| Local street | -0.765*** (-0.792, -0.738) | 15.235*** (14.113, 16.358) | 181.112*** (173.496, 188.728) |
| 2-3 lanes | 0.085*** (0.073, 0.097) | -3.241*** (-3.806, -2.676) | -13.913*** (-17.742, -10.083) |
| >3 lanes | 0.059*** (0.039, 0.079) | -7.068*** (-8.130, -6.006) | -53.100*** (-60.297, -45.904) |
| Legal divider | 0.045 (-0.037, 0.128) | 0.662 (-1.437, 2.761) | -6.457 (-20.700, 7.787) |
| Physical divider | -0.025 (-0.064, 0.014) | 1.630. (-0.122, 3.383) | -1.675 (-13.567, 10.217) |
| Intersection | -0.090*** (-0.101, -0.080) | 1.129*** (0.668, 1.590) | 2.977. (-0.154, 6.107) |
| Frontage | 0.124*** (0.107, 0.141) | -0.931* (-1.748, -0.114) | -34.517*** (-40.064, -28.970) |
| Min. altitude | -0.149*** (-0.161, -0.137) | 2.894*** (2.329, 3.460) | 42.726*** (38.925, 46.527) |
| Slope | -0.014** (-0.023, -0.006) | 0.213 (-0.196, 0.623) | 3.920** (1.142, 6.698) |
| Flood severity | 0.021*** (0.014, 0.027) | -1.620*** (-2.026, -1.213) | -22.151*** (-24.910, -19.392) |
| **Smooth terms (Estimated degrees of freedom (e.d.f.))** | | | |
| ti (Latitude, Longitude) | 13.400*** | 14.950*** | 13.326*** |
| **Model fit** | | | |
| R2 (Adjusted) | 0.280 | 0.204 | 0.381 |
| Deviance explained | 28.2% | 20.5% | 38.1% |
| **Winter Storm (February 2 - 5, 2022)** | | | |
| Variables | Duration | Speed Change | AUC |
| (Intercept) | 3.738*** (3.729, 3.746) | -26.864*** (-27.763, -25.964) | -1263.003*** (-1285.813, -1240.193) |
| **Linear effects (Parametric coefficients)** | | | |
| Arterial | -0.247*** (-0.257, -0.238) | 1.647** (0.607, 2.688) | 486.977*** (460.626, 513.328) |
| Collector | -0.532*** (-0.539, -0.525) | 7.308*** (6.532, 8.085) | 779.141*** (759.494, 798.789) |
| Local street | -0.842*** (-0.858, -0.825) | 13.911*** (12.529, 15.293) | 961.625*** (926.607, 996.644) |
| 2-3 lanes | 0.191*** (0.183, 0.198) | -5.913*** (-6.608, -5.218) | -170.444*** (-188.071, -152.818) |
| >3 lanes | 0.035*** (0.023, 0.048) | 1.535* (0.230, 2.841) | -73.913*** (-107.025, -40.802) |
| Legal divider | -0.015 (-0.044, 0.014) | 1.444 (-1.140, 4.029) | 45.915 (-19.656, 111.486) |
| Physical divider | -0.131*** (-0.156, -0.106) | 3.815*** (1.657, 5.973) | 63.217* (8.479, 117.955) |
| Intersection | -0.246*** (-0.252, -0.239) | 1.496*** (0.928, 2.064) | 144.641*** (130.232, 159.049) |
| Frontage | 0.105*** (0.095, 0.116) | -1.503** (-2.509, -0.496) | -108.675*** (-134.204, -83.147) |
| Min. altitude | -0.006. (-0.013, 0.001) | -0.895* (-1.585, -0.204) | -4.685 (-21.157, 11.787) |
| Slope | -0.000 (-0.006, 0.005) | 0.186 (-0.318, 0.690) | -8.453 (-21.212, 4.307) |
| Snow severity | -0.014*** (-0.018, -0.009) | 0.314 (-0.186, 0.814) | 5.903 (-6.778, 18.584) |
| **Smooth terms (Estimated degrees of freedom (e.d.f.))** | | | |
| ti (Latitude, Longitude) | 15.801*** | 13.582*** | 11.248*** |
| **Model fit** | | | |
| R2 (Adjusted) | 0.263 | 0.121 | 0.397 |
| Deviance explained | 25.8% | 12.2% | 39.8% |

Note: Robust 95% confidence interval (CI) is in parentheses. Significance codes: 0 '***' 0.001 '**' 0.01 '*' 0.05 '.' 0.1 " 1. P-value <0.05 is considered statistically significant. ti () means the marginal nonlinear interaction function. Coefficients are standardized to facilitate comparisons among models and variables.

Among the linear effects, the coefficients for speed change and AUC share the same sign, while the coefficients for duration have the opposite sign. This is plausible since more negative speed change, more negative AUC, and longer event duration, all indicate more severe impacts of weather events on road traffic. However, the significance of these coefficients may not be consistent across the three measures, indicating nuances do exist when measuring resilience from different aspects. In the following section, we will discuss the coefficients categorized by independent variables.

*Functional Class:* Among all variables, the functional class presents the strongest relationship with traffic resilience. Controlling for other variables, freeways experience the longest recovery time and the greatest speed reduction, followed by arterials, collectors, and local streets. Compared to freeways, local streets experience 13.73% to 15.21% less speed reduction and require 53.46% ($1 - e^{-0.765}$) to 56.91% ($1 - e^{-0.842}$) less time for recovery. It is plausible since higher-class roads carry more traffic volume and are more prone to congestion when their capacity is curtailed. These findings also align with similar studies conducted in other cities under diverse weather conditions (Malin et al., 2019; Praharaj et al., 2021a).

*Lane Category:* During winter storms, roads with 2-3 lanes experience the longest time to recover and undergo the greatest reduction in speed compared to roads with other lane numbers. This observation may be attributed to the fact that roads with 2-3 lanes often fall within a middle range in terms of capacity and importance. On the one hand, they receive relatively fewer resources and attention in recovery efforts, such as debris clearance and damage repair, when compared to higher-class roads with more lanes. On the other hand, compared to local streets with only one lane, roads with 2-3 lanes may endure greater traffic pressure shifting from affected higher-class roads whose capacity is substantially reduced during extreme weather.

*Divider Type, Intersection, and Frontage:* The presence of dividers is not significantly associated with traffic resilience under floods. However, during winter storms, roads with physical dividers tend to recover faster and experience less speed reduction compared to roads without dividers. This may be due to snowplows being able to clear the lanes more effectively without pushing snow into opposing directions on roads with physical dividers. In addition, intersection links show faster recovery and less speed reduction. One explanation is that intersections are mainly located on local streets that are less impacted by weather events. Last, frontage roads experience greater speed reduction and take longer to recover compared to others.

Frontage roads run parallel to high-class freeways, making them vulnerable to higher traffic pressure if the nearby high-class freeways are blocked during weather events.

*Geometry:* Road geometry, including slope and altitude, is significantly related to road traffic resilience under floods; however, it does not show significance when modeling traffic resilience under winter storms. During floods, roads with higher altitudes and steeper slopes tend to recover faster and experience less speed reduction, which is reasonable since higher altitude reduces the risk of inundation, while steeper slopes facilitate faster flood dissipation.

*User-perceived severity*: Roads with higher user-perceived severity exhibit longer recovery time and greater speed reduction during floods. This, to some extent, demonstrates the ability of crowdsourced data in capturing road link-level impact during floods. However, the ability disappears under winter storms. User-perceived severity does not emerge as a significant predictor of traffic speed change during winter storms, and even a counterintuitive negative relationship is observed between the link severity and its recovery time. This is very likely because, during more severe events like winter storms, very few users can drive on the road to serve as the "sensors", especially on those fully-blocked road segments.

**4.4 Relationship between traffic resilience and normal traffic flow characteristics**

Due to the high multicollinearity between road characteristics and traffic flow characteristics, we separately fit the GAM for traffic flow characteristics. Results show high e.d.f. (>1) for their relationships, indicating the presence of high nonlinearity, as graphically illustrated in **Figure 8**. Some key insights can be concluded as follows. First, average speed and AADT exhibit broad positive associations with event duration while being negatively associated with speed change and AUC. Alternatively, road links characterized by higher average speeds and greater AADT during normal periods experience more pronounced impacts and need longer time to recover when confronted with adverse weather events. This is in line with the findings in **Table 4** as links with higher speeds and greater AADT are more likely to belong to higher-class roads. Second, the complexity of the relationships is highlighted by the discernible presence of non-monotonic patterns and threshold effects (Hu et al., 2023a). The gradients of relationships are high in their initial stages, followed by a gradual leveling off as the relevant independent variable continues to ascend, with some of them even exhibiting a sign reversal. For example, a slightly rebounding pattern is observed in the relationship between AUC and traffic flow characteristics within the tail of each panel. This indicates impacts suffered by roads with the highest average speed and AADT are not inherently greater than that sustained by slightly lower-ranking counterparts. This pattern echoes the trends observed in lane category relations in **Table 4** and can be explained by similar reasons, i.e., top-class roads have secured foremost attention by local agencies in the face of adverse weather events, thereby expediting their recovery processes.

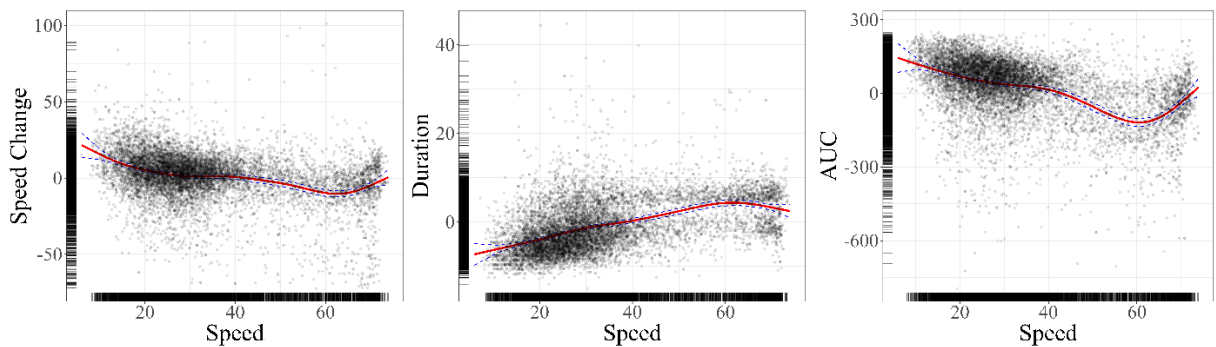

(a) Relationships with average speed during flood (August 21 - 23, 2022)

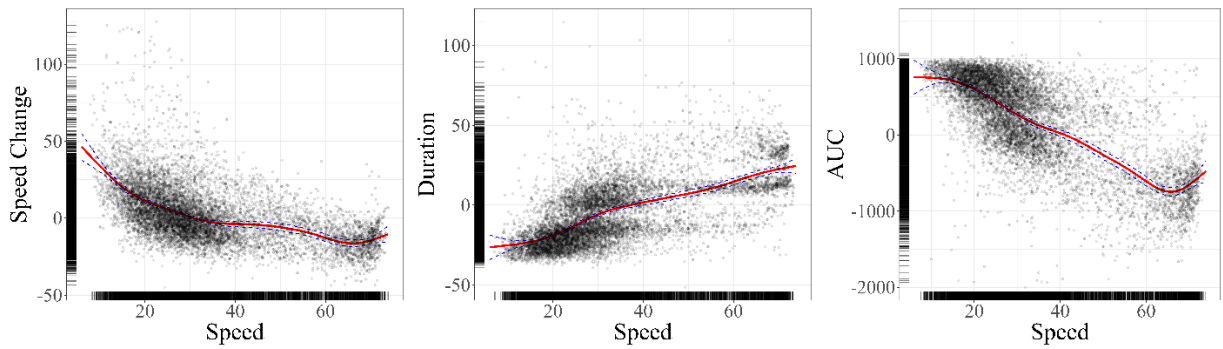
(b) Relationships with average speed during winter storm (February 2 - 5, 2022)

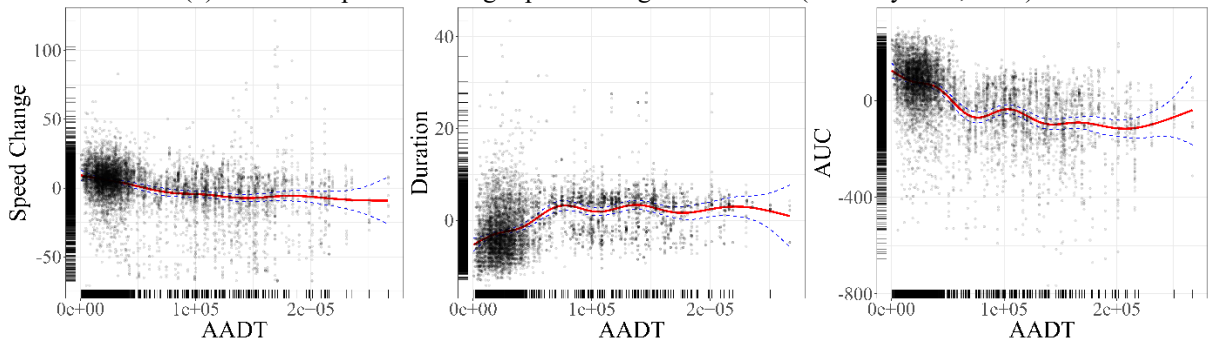
(c) Relationships with AADT during flood (August 21 - 23, 2022)

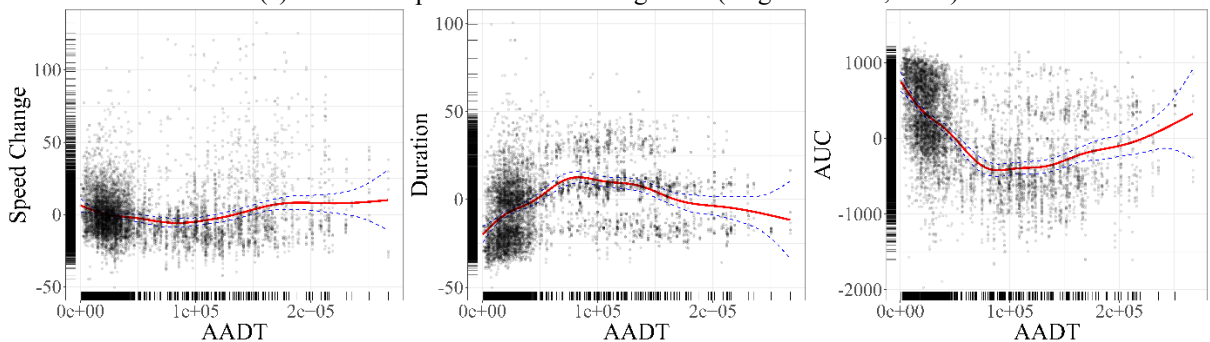
(d) Relationships with AADT during winter storm (February 2 - 5, 2022)

**Figure 8**. Nonlinear relationship between traffic flow characteristics and three resilience metrics.

The top two rows represent the August flood while the bottom two rows represent the February first winter storm. For each panel, the y-axis is one resilience metric estimated by the independent variable shown on the x-axis via the GAM. The red curve is the average trend. Blue dashes are 95% CI. Each black spot represents one road link.

## 5. Policy implications

Given that road functional class has the strongest relationship with traffic resilience, it is crucial for transportation authorities to prioritize road maintenance and recovery efforts based on the road functional class. Freeways, being the most critical for carrying high-mobility large-volume vehicle movements, should receive the highest priority in terms of resources and attention during weather-related disruptions, followed by arterials, collectors, and local roads. This loose prioritization approach optimizes the allocation of limited maintenance resources efficiently. However, a judicious equilibrium should be maintained to avoid disproportionately channeling excessive resources toward the highest-level roads. Using the functional class of the road as a guide, leveraging a mix of big data sources to track and predict link-level traffic changes in near-real-time, and adjusting resources accordingly may be the optimal solution.

While functional class is the primary rule for resource allocation, other road characteristics should also be considered. Roads with 2-3 lanes tend to experience longer recovery time and greater speed reduction during extreme weather. Hence, transportation authorities should allocate additional resources and attention to these roads during post-event recovery efforts, especially if they also belong to high functional classes, exhibit high average speed and AADT, or serve as frontage roads along affected freeways. Additionally, roads with lower altitudes and flatter slopes are more vulnerable during floods. Road geometry should be considered when planning and designing road infrastructure to improve flood resilience, particularly in low-lying and flat areas.

Although winter storms are less frequent in the DFW area, their impacts on road traffic are more pronounced than floods. This indicates in the DFW area, while engineering and infrastructure projects have already mitigated flood risks to a certain extent, winter storm management has not received the same level of attention and investment. Additional efforts are needed to make this region well-prepared for winter extreme weather. This includes conducting regular preventive maintenance on roads and bridges, implementing early warning systems, establishing a robust snow removal and plowing strategy, deploying salt and sand strategically on road surfaces, and encouraging the use of alternative transportation options.

Crowdsourced data demonstrate their ability to monitor road traffic changes and capture travelers' responses during unexpected weather events. Their high spatiotemporal resolution provides valuable insights to transportation authorities to implement targeted and nuanced recovery measures. However, the trustworthiness of crowdsourced data deserve further discussion. On the one hand, under moderate weather conditions, crowdsourced reports may increase in number, but not all of them can be deemed reliable. Factors such as the lack of expertise, sampling biases, and selective reporting may lead to inaccuracies in the data. Studies have shown that only 71.7% of flooding reports from Waze are predicted to be trustworthy (Praharaj et al., 2021b). On the other hand, under more severe weather conditions, crowdsourced data may lose its ability to reflect real-time road conditions due to limited user participation. For instance, if heavy snow freezes the entire road network, there might be no active drivers on the roads to report information. Hence, decision-makers must be cautious in interpreting and using crowdsourced data. Combining crowdsourced data with alternative sources, such as roadside detectors, traffic surveillance systems, or remote sensing technologies, can enhance data accuracy and completeness during various weather events, ultimately aiding in the implementation of effective mitigation and recovery strategies.

## 6. Conclusions and Limitations

This study comprehensively investigates the impact of various weather events on road traffic resilience using crowdsourced data. The findings reveal that winter storms have the most significant impact on traffic, followed by floods, while the effects of fog are relatively minor. In addition, high-class roads with greater average speed and volume bear a greater reduction in speed and volume and take longer to recover. Other link characteristics, such as lane number and link geometry, also significantly influence road resilience. This study highlights the effectiveness of crowdsourced data in monitoring road traffic changes during weather events, outperforming traditional agency sources by its extensive spatiotemporal coverage, fine resolution, and real-time monitoring capabilities. However, the study also emphasizes the need for caution in using crowdsourced data, as some reports may lack reliability, particularly under severe weather conditions. By combining crowdsourced data with other sources, decision-makers can better navigate the challenges posed by diverse weather conditions and implement more effective measures to ensure road resilience.

Several limitations are recognized and deserve further research. First, relationships explored in our models are correlational and do not establish causality. Certain road characteristics, such as the correlation between lane numbers and functional classes, may confound the interpretation of causal relationships. Second, only the speed data are used in this study to assess link-level road traffic resilience. Additional traffic flow characteristics, such as volume and density, can be incorporated to evaluate traffic resilience more comprehensively if related data are available. Last, only one region is analyzed in this study and results may be region-specific. Caution should be exercised when extrapolating the findings to other regions with different weather and road network patterns. The authors are also working on the comparison among different regions to enhance the generalizability of the above findings.

Qiang, Y., Xu, J., 2020. Empirical assessment of road network resilience in natural hazards using crowdsourced traffic data. *International Journal of Geographical Information Science* 34(12), 2434-2450.

Rahman, R., Hasan, S., 2023. A deep learning approach for network-wide dynamic traffic prediction during hurricane evacuation. *Transportation Research Part C: Emerging Technologies* 152, 104126.

Shen, S., Kim, K., 2020. Assessment of transportation system vulnerabilities to tidal flooding in Honolulu, Hawaii. *Transportation research record* 2674(11), 207-219.

Singh, P., Sinha, V.S.P., Vijhani, A., Pahuja, N., 2018. Vulnerability assessment of urban road network from urban flood. *International journal of disaster risk reduction* 28, 237-250.

Stott, P., 2016. How climate change affects extreme weather events. *Science* 352(6293), 1517-1518.

Theofilatos, A., Yannis, G., 2014. A review of the effect of traffic and weather characteristics on road safety. *Accident Analysis & Prevention* 72, 244-256.

TRB, T.R.B., 2010. *HCM2010*.

Wang, H.-W., Peng, Z.-R., Wang, D., Meng, Y., Wu, T., Sun, W., Lu, Q.-C., 2020a. Evaluation and prediction of transportation resilience under extreme weather events: A diffusion graph convolutional approach. *Transportation research part C: emerging technologies* 115, 102619.

Wang, T., Hu, S., Jiang, Y., 2020b. Predicting shared-car use and examining nonlinear effects using gradient boosting regression trees. *International Journal of Sustainable Transportation*, 1-15.

Wood, S.N., 2003. Thin plate regression splines. *Journal of the Royal Statistical Society: Series B (Statistical Methodology)* 65(1), 95-114.

Xing, J., Wu, W., Cheng, Q., Liu, R., 2022. Traffic state estimation of urban road networks by multi-source data fusion: Review and new insights. *Physica A: Statistical Mechanics and its Applications* 595, 127079.

Yabe, T., Jones, N.K., Rao, P.S.C., Gonzalez, M.C., Ukkusuri, S.V., 2022. Mobile phone location data for disasters: A review from natural hazards and epidemics. *Computers, Environment and Urban Systems* 94, 101777.

Yuan, F., Lee, C.-C., Mobley, W., Farahmand, H., Xu, Y., Blessing, R., Dong, S., Mostafavi, A., Brody, S.D., 2023. Predicting road flooding risk with crowdsourced reports and fine-grained traffic data. *Computational Urban Science* 3(1), 15.

Zhang, N., Alipour, A., 2020. Multi-scale robustness model for highway networks under flood events. *Transportation research part D: transport and environment* 83, 102281.

Zhang, N., Alipour, A., 2021. A multi-step assessment framework for optimization of flood mitigation strategies in transportation networks. *International Journal of Disaster Risk Reduction* 63, 102439.

Zhao, T., Zhang, Y., 2020. Transportation infrastructure restoration optimization considering mobility and accessibility in resilience measures. *Transportation Research Part C: Emerging Technologies* 117, 102700.

Zhou, M., Wang, D., Li, Q., Yue, Y., Tu, W., Cao, R., 2017. Impacts of weather on public transport ridership: Results from mining data from different sources. *Transportation research part C: emerging technologies* 75, 17-29.
36